\documentstyle[12pt]{article}
\begin{document}
\def\thefootnote{\fnsymbol{footnote}}
\begin{flushright}
KANAZAWA-94-06\\
March, 1994
\end{flushright}
\vspace{13mm}
\begin{center}
{\Large\bf Gauge Coupling Unification due to Non-universal Soft
Supersymmetry Breaking}
\vspace{1.5cm}\\
{\large Tatsuo Kobayashi
\footnote[1]{e-mail:kobayasi@hep.s.kanazawa-u.ac.jp},
 Daijiro Suematsu
\footnote[2]{Work supported in part by a Grant-in-Aid for Scientific
Research from the Ministry of Education, Science and Culture(\#05640337)
}
\footnote[3]{e-mail:suematsu@hep.s.kanazawa-u.ac.jp}\\
 and\\
 Yoshio Yamagishi
\footnote[4]{e-mail:yamagisi@hep.s.kanazawa-u.ac.jp}
}
\vspace{1cm}\\
{\it Department of Physics, Kanazawa University\\
 Kanazawa 920-11, JAPAN}
\vspace{1.5cm}
\end{center}
{\large\bf Abstract}\\
Gauge coupling unification is studied in the MSSM with non-universal 
soft supersymmetry breaking terms.
If gaugino masses are sufficiently smaller than scalar soft masses
and the scalar soft masses have also certain types of non-universality,
gauge coupling unification scale can be larger than 
$3\times 10^{16}$~GeV even within the MSSM contents.
String unification may not need a large threshold correction 
or a large modulus value.
We also discuss the relation to the string model building.
\newpage
\setcounter{footnote}{0}
\def\thefootnote{\arabic{footnote}}
Superstring theory is the presently known unique theory which unifies 
all interactions including the gravity.
Various features of the superstring unification are studied by now.
The unification of the gauge coupling constants is one of the expected 
features.
Its unification is different from the usual grand unification scenario
and do not need a unification group like $SU(5)$ or $SO(10)$.
The gauge coupling unification 
$k_3g^2_3=k_2g^2_2=k_1g^2_1$
is brought due to the fact that all gauge interactions are induced 
from the affine Kac-Moody algebras on the world sheet\cite{G}.
Its unification scale is estimated as 
$M_{\rm str} \simeq 0.5\times g_{\rm str}\times 10^{18}$~GeV
\cite{K}.
Recent study based on the precise measurements at LEP shows that 
the gauge coupling constants of $SU(3)\times SU(2)\times U(1)$ 
correctly meet at $M_X \simeq 3\times 10^{16}$~GeV in the minimal 
supersymmetric standard model(MSSM)\cite{LP}.
The explanation of this discrepancy between $M_{\rm str}$ and $M_X$ 
is an important issue for building up superstring inspired models.

Some stringy explanations for the discrepancy are proposed by now. 
One of such possibilities is based on the existence of additional 
massless fields which become massive at an intermediate 
scale\cite{AEKN}.
In general there are extra massless colored modes beyond the MSSM 
spectrum in the superstring models.
However, the inclusion of these fields usually causes various 
phenomenological problems like the proton decay.
Moreover there are too many degrees of freedom to make some predictions.
Thus as the first trial it seems more promising to find other 
explanation which works within the MSSM spectrum at least 
for the unification of $SU(3)$ and $SU(2)$ factor groups.
In superstring there are infinite number of massive modes around 
$M_{\rm pl}$.
These modes may bring large threshold corrections to the gauge 
coupling constants at $M_{\rm str}$\cite{DKL}.
If this is the case, the gauge coupling constants split at 
$M_{\rm str}$ and appear to coincide at $M_X$.
This possibility has been studied using the MSSM spectrum at the low 
energy region[6-10].
The models are stringently constrained to realize this scenario.
Every field of the MSSM must have the nontrivial modular weight and 
also an overall modulus is required to have a large vacuum 
expectation value.
For example, the $Z_N$ ($Z_N \times Z_M$) orbifold models require 
$Re T \equiv T_R\geq 7$ ($3$) to obtain the large threshold corrections 
consistent with the measured values of the coupling at $M_{\rm Z}$.

As is well-known, the superstring theory generally has the stringy 
symmetry called the target space modular invariance\cite{KY}.
If we impose this invariance on the model, 
the potential minimum will be around the selfdual point 
$T_R \sim \sqrt 2$.
In this case we can not expect the large threshold correction.
From this viewpoint, it is very interesting to study the 
possibility of pulling up the unification scale $M_X$ to near 
$M_{\rm str}$ without the large threshold correction in the MSSM 
spectrum.
In this letter we investigate this point noting non-universal
soft supersymmetry breaking masses.

First of all we briefly review the string threshold correction\cite{DKL} 
and the soft supersymmetry breaking masses\cite{IL,BIM,KV}.
In the following we concentrate ourselves to the case with an 
overall modulus.
The generalization will be done in the straightforward way.
As mentioned above, it is expected that superstring theory is 
invariant under the following target space modular 
transformation\cite{FLST}:
\begin{eqnarray}
&&T \rightarrow {aT-ib \over icT+d} ; \quad ad-bc=1 , \quad 
a, b, c, d \in {\bf Z},\\
&&C_i \rightarrow (icT+d)^{n_i}C_i,
\end{eqnarray}
where $T$ is an overall modulus and $C_i$ is a matter field.
A modular weight $n_i$ is an integer. 
Imposing the target space modular invariance, the threshold correction 
to the gauge coupling constants is calculated in certain types of
 orbifold models.
Due to such an effect the gauge coupling constants at $M_{\rm str}$ 
are effectively shifted as\cite{DKL}
\begin{equation}
{1 \over g_a^2}={k_a \over g_{\rm str}^2} - {1 \over 16\pi^2}
(b_a^\prime -k_a\delta_{GS})\log[(T+T^\ast)\vert\eta(T)\vert^4],
\end{equation}
where $k_a$ is the Kac-Moody level of the gauge group $G_a$ and
$\delta_{GS}$ is a gauge group independent constant.
It cancels a part of the duality anomaly in the same way as 
Green-Schwarz mechanism of the $U(1)$ gauge\cite{DSKZ}.
Dedekind function $\eta$ is expressed as
\begin{equation}
\eta(T)=e^{-\pi T/12}\prod^\infty_{n=1}(1-e^{-2\pi nT}).
\end{equation}
A duality anomaly coefficient $b_a^\prime$ is related to a coefficient 
$b_a$ of a one-loop $\beta$-function in the MSSM as
\begin{equation}
b_a^\prime=b_a +2\sum_i T_a(C_i)(1+n_i).
\end{equation}
Here $T_a(C_i)$ is a second order index of the field $C_i$ for $G_a$.
The gauge coupling unification has been examined based on these formulae 
and the universal soft supersymmetry breaking terms within the MSSM
framework.
It is also suggested that the unification at $M_{\rm str}$ is 
possible
if each superfield in the MSSM has a certain modular weight and the 
value of $T_R$ is rather large.
However, in the duality invariant theory the minimum of the potential 
is realized around the selfdual point $T_R\sim \sqrt 2$.
Actually, in the gaugino condensation scenario the selfdual point 
appears as the potential minimum\cite{FILQ}.\footnote{Recently 
it is suggested that the large $T_R$ can be possible
if we consider the loop correction in the gaugino condensation 
mechanism\cite{MR}.}
If this is the case, there seems to be the contradiction.

As is known from the study of the soft supersymmetry breaking terms,
the fields with different modular weights have the non-universal soft 
masses.
In the orbifold models with zero cosmological constant the scalar 
masses $m_i$ and gaugino masses $M_a$ at $M_{\rm str}$ are 
represented as\cite{IL,BIM}
\begin{eqnarray}
&&m_i^2=m_{3/2}^2\left(1+n_i\cos^2\theta\right)\\
&&M_a=\sqrt 3m_{3/2}\left({k_a {\rm Re}S\over {\rm Re}f_a}
\sin\theta+
\left({(b^\prime_a-k_a\delta_{\rm GS}(T+T^\ast)\hat G_2(T+T^\ast)\over 
32\pi^3\sqrt 3{\rm Re}f_a}\right)\cos\theta \right)
\end{eqnarray}
where $S$ is a dilaton field and $m_{3/2}$ is a gravitino mass.
$f_a$ is a gauge kinetic function of $G_a$.
The nonholomorphic Eisenstein function $\hat G_2$ is defined as
$$
\hat G_2\left(T+T^\ast\right) =-4\pi\left(\partial \eta(T)/\partial T)
(\eta(T)\right)^{-1}-2\pi /(T+T^\ast).
$$
A goldstino angle $\theta$ expresses the feature of 
the supersymmetry breaking.
This fact suggests that we should carefully treat the threshold 
correction due to the non-universal soft masses at the low energy region 
in the renormalization group study, especially if we consider the 
models with nontrivial modular weights.

At the low energy region the soft supersymmetry breaking masses
are determined by the following supersymmetric one-loop renormalization 
group equations\footnote{If some superpartners decouple at $M_S$, the 
one loop $\beta$-function coefficient $b_a$ in the eq.(9) happens to 
be modified below $M_S$. 
It is also different from $\bar b_a$ in eq.(12).
This is because one-loop corrections to the gaugino mass include
 graphs which have both of the fermions and their superpartners 
simultaneously in internal lines.
In the following analysis we take account of this point.}
\begin{eqnarray}
&&{dm_i^2 \over dt}={1 \over 8\pi^2}\left(-4\sum_a C_a(i)M_a^2g_a^2 
+({\rm Yukawa\ terms})\right),\\
&&{dM_a \over dt}={b_a \over 8\pi^2}g_a^2M_a,
\end{eqnarray}
where $C_a(i)$ is the quadratic Casimirs for each scalar labeled 
by $i$.
If we neglect Yukawa coupling contributions
\footnote{Except for the contribution of top Yukawa coupling, this 
approximation will be completely justified.
We will return to this point later.},
 these equations can be easily solved analytically and results are
\begin{eqnarray}
&&m_i^2(Q)=m_i^2(M_{\rm str}) +\sum_a{2C_a(i) \over b_a}
\left(1-{1 \over (1+b_a{g^2(M_{\rm str}) \over 8\pi^2}
\ln{M_{\rm str} \over Q})^2}\right)
M_a^2(M_{\rm str}),\\
&&M_a(Q)={M_a(M_{\rm str}) \over g^2(M_{\rm str})}
\left({g^2(M_{\rm str}) \over 
1+b_a{g^2(M_{\rm str}) \over 8\pi^2}\ln{M_{\rm str} \over Q}} 
\right).
\end{eqnarray}
If these masses widely split, their threshold corrections can affect
the evolution of the gauge coupling constants and then the unification 
scale.
Hereafter noting this point, we study the relation between the 
unification scale and the soft breaking masses.
In the following study we consider the unification of $SU(3)$ and 
$SU(2)$ alone because the Kac-Moody level of $U(1)$ is a free 
parameter in the superstring theory\cite{Ib}.

We now classify the models by the mass patterns of the gauginos, 
squarks and sleptons at the low energy region.
To simplify the analysis, we divide the fields of the MSSM into two 
groups named as $A$ and $B$.
Group $A$ is a set of superpartners which decouple from the 
renormalization group equations at $M_S(\ge M_Z)$.
The remaining superpartners belong to Group $B$ and contribute to them 
down to $M_Z$.
This procedure will be sufficient to see the qualitative feature of the 
gauge coupling unification.
As we only consider the unification scale of $SU(3)$ and $SU(2)$,
the relevant superpartners in the MSSM are squark doublets $Q$, squark 
singlets $U,~D$, slepton doublets $L$,\footnote{Later we shall discuss
the $U(1)$ gauge coupling where the slepton singlets will be treated 
in the similar way.} Higgsino $H_1,~H_2$ and gauginos 
$\lambda_3,~\lambda_2$.
The typical cases presented here are followings:\\
\begin{tabular}{l}
{\it Case~I~{\rm (ordinary MSSM)}}: \quad
$A=\{Q,~U,~D,~L,~H_1,~H_2,~\lambda_3,~\lambda_2\}$,\\
{\it Case~II}: \quad
$A=\{Q,~U,~D,~L,~H_1,~H_2\},\ B=\{\lambda_3,~\lambda_2\}$,\\
{\it Case~III}: \quad
$A=\{Q,~L,~H_1,~H_2\},\ B=\{U,~D,~\lambda_3,~\lambda_2\}$,\\
{\it Case~IV}: \quad
$A=\{L,~H_1,~H_2\},\ B=\{Q,~U,~D,~\lambda_3,~\lambda_2\}$,\\
{\it Case~V}: \quad
$A=\{Q,~U,~D,~H_1,~H_2 \},\ B=\{L,~\lambda_3,~\lambda_2\}$,\\
{\it Case~VI}: \quad
$A=\{U,~D,~H_1,~H_2\},\ B=\{Q,~L,~\lambda_3,~\lambda_2\}$,\\
\end{tabular}\\
where the generation indices are abbreviated.
For the Higgs scalars we confine ourselves to the situation in each case
that one Higgs doublet decouples at $M_S$.

Here we should note some points on the non-universality of soft scalar 
masses.
At first to justify the above classification the gaugino masses 
should not be large so as not to erase the difference in the soft 
scalar masses.
Otherwise, as seen from the renormalization group equations of 
the scalar masses, the contribution to the scalar mass from 
gauginos becomes dominant and erases the non-universality at the low 
energy region.
Secondly the non-universal soft squark masses are dangerous for the 
flavor changing neutral currents. 
To avoid it we need to choose the non-universality which induces
no dangerous mass difference between the generations.
Finally $M_S$ can not be so large from the naturalness argument.
It should be at most a few TeV.

Now we study the unification scale $M_X$ of $SU(3)$ and $SU(2)$ gauge 
couplings beginning from the low energy region.
They are related by the renormalization group equation as
\begin{equation}
\alpha_a^{-1}(M_X)=\alpha_a^{-1}(M_S)
-{b_a \over 2\pi}\ln{M_X \over M_S}, \qquad
\alpha_a^{-1}(M_Z)=\alpha_a^{-1}(M_S)
-{\bar b_a \over 2\pi}\ln{M_Z \over M_S}.
\end{equation}
Using these formulae, we found that the unification scale $M_X$ is 
expressed as the function of $M_S$,
\begin{equation}
M_X=M_S({M_Z \over M_S})^{\bar b_3-\bar b_2 \over b_3-b_2}
\exp\left({2\pi \over b_3-b_2}
\left(\alpha^{-1}_3(M_Z)-\alpha^{-1}_2(M_Z)\right)\right).
\label{unis}
\end{equation}
Our models are completely equivalent to the MSSM from $M_X$ to $M_S$ 
so that $b_3=-3$ and $b_2=1$.
In the region below $M_S$ $\bar b_a$ is different in each case.
The values of $(\bar b_3, \bar b_2)$ are followings:
$(-7, -19/6)$ in $Case~I$, $(-5, -11/6)$ in $Case~II$, $(-4, -11/6)$ 
in $Case~III$,
$(-3, -1/3)$ in $Case~IV$, $(-5, -4/3)$ in $Case~V$, $(-4, 1/6)$
 in $Case~VI$.
As easily seen from eq.(\ref{unis}), the smaller value of 
$(\bar b_3- \bar b_2)/(b_3-b_2)$ is preferable for our scenario.
The unification scale $M_X$ can be estimated for the various values 
of $M_S$ if we use 
\begin{equation}
M_Z=91.173~GeV,\ 
\alpha^{-1}(M_Z)=127.9,\ 
\sin^2\theta_W(M_Z)=0.2328,\
\alpha_3(M_Z)=0.118,
\end{equation}
as the input data\cite{par}.
Figure~1 shows the change of the unification scale $M_X$ against the 
decoupling scale $M_S$ of the superpartners in Group A for each case. 
It is remarkable that $M_X$ becomes larger accompanied with the increase
of $M_S$ in {\it Case~II $\sim$ V}.
This feature is very different from {\it Case~I} in which 
$M_X \sim 3\times 10^{16}$~GeV is almost stable against $M_S$.
It should be also noted that the unification scale $M_X$ moves upward
if the squark doublet decouples at $M_S$.
The non-universal soft masses tend to give the higher
unification scale than the case of the universal soft masses.
If this qualitative tendency is the case, $M_X$ can reach 
$M_{\rm str}$ 
even if the threshold correction is not so large.
From the quantitative point of view we should note that there is 
an ambiguity of order $10^{0.3}$~GeV also in the present estimation 
of $M_X$ as usual. 

Using eqs.(3) and (12), the necessary threshold correction is 
estimated as
\begin{equation}
\sqrt{T+T^\ast}\vert\eta(T)\vert^2=
({M_{\rm str}\over M_X})^{b_3-b_2 \over b_3^\prime -b_2^\prime},
\end{equation}
where we take the Kac-Moody level as $k_3=k_2=1$.
As an example, let us adopt {\it Case~III} and estimate the  
threshold correction required to realize the unification at 
$M_{\rm str}$.
In ref.\cite{KKO2} it is shown that the MSSM derived from the $Z_N$ 
orbifold models can have $b^\prime_3-b^\prime_2=3$ or 4 and for 
$Z_6$-II the maximum value of $b^\prime_3-b^\prime_2$ is equal to 6.
Putting $M_S=1$~TeV and 3~TeV, we have $M_X=10^{17.0}$~GeV 
and $10^{17.2}$~GeV, respectively(see Fig.1).
In the case where $M_X=10^{17.0}$GeV and $b^\prime_3-b^\prime_2=3, 4$ 
and 6, we obtain $T_R=5.5, 4.5$ and 3.5, respectively, using (15) and 
$M_{\rm str}=3.7 \times 10^{17}$GeV.
These values of $T_R$ are fairly smaller than ones estimated in the 
universal soft breaking case where the unification scale is 
$M_X=10^{16.5}$~GeV.
For example, in the case of $M_X=10^{16.5}$GeV the difference 
$b^\prime_3-b^\prime_2=3$ leads to $T_R=9$.
Further in the case where $M_X=10^{17.2}$GeV and 
$b^\prime_3-b^\prime_2=3$ and 4, we can have $T_R=4$ and 3.5.
The $Z_N \times Z_M$ orbifold models can have larger values of 
$b^\prime_3-b^\prime_2$ than the $Z_N$ orbifold models \cite{KKO} 
and then derive the smaller values of  $T_R$, e.g. $T_R<2$ in the case 
of $M_X=10^{17}$~GeV.

Next we shall consider in what type of superstring models 
the favorable soft breaking masses presented in the previous part
are realized.
From the recent study of the soft supersymmetry breaking terms, 
we know their general features at $M_X$\cite{BIM}.
On the other hand, we can transmute the soft masses at the low energy 
region into the ones at $M_X$ using eqs.(10) and (11) in our present 
cases.
Comparing them we can know what kinds of minimal superstring 
standard models need not 
the large threshold correction for the string unification.
We show the change of the soft masses from the low energy region to
$M_X$ against the gaugino mass $M_U$ at $M_X$ for each case in Table~1.
As mentioned in the previous part the large gaugino mass will dilute
the non-universality in the soft scalar masses by the renormalization 
group effect.
This imposes a certain condition on the upper bound of the gaugino mass
to make our scheme work.
We can find from Table~1 that the dilution effects of the 
non-universality will be escapable if $m_i^2(M_X)/M_U^2>10$ is satisfied.

It is very interesting to know in what type of supersymmetry breaking
this situation is generally realized.
As discussed in ref.\cite{BIM} such soft terms can be caused 
in the moduli dominated supersymmetry 
breaking (large $\cos^2\theta$). 
However, our scenario needs various modular weights $n_i \le -2$
for the non-universality in $Case III \sim VI$.
The goldstino angle $\cos^2\theta$ can not be so large to 
guarantee $m_i^2(M_X)>0$ 
because of its modular weight dependence as seen from eq.(6).
In such cases generally the dilaton contribution to the soft breaking 
masses is dominated and then $m_i^2/M_U^2<1$ for the suitable values 
of $\cos\theta$ and $T_R$ at $M_X$.
The original non-universality in the soft scalar masses may be diluted 
away. The more careful study for these cases will be necessary .
The most promising case where $m_i(M_X)/M_U>1$ is {\it Case~II}.
Such soft masses can be easily realized in
the orbifold model in which the modular weights of all massless modes 
are $n_i=-1$ and also the gaugino condensation model
as suggested in ref.\cite{CCM}
, where $M_S$ is estimated as $M_S=1 \sim 4$~TeV.

Some comments are in order.
Firstly 
we have introduced the soft scalar masses which are degenerate 
between the different generations in the same type flavors.
The non-universality presented here will not yield the dangerous FCNC.
Secondly 
we neglected the Yukawa couplings in the renormalization 
group equations to estimate the scalar masses.
Except for the case of the top sector this treatment will be justified.
The top Yukawa reduces the stop mass at the low energy region.
The stop mass at $M_X$ must be large enough to keep the degeneracy 
between the same flavor at $M_Z$.
Anyway its effect will not affect our results crucially.
Thirdly 
we do not refer to the unification of $U(1)$.
However, its occurrence can be expected by choosing a suitable value 
of the Kac-Moody level $k_1$ as suggested in ref.\cite{Ib,KKO,KKO2}.
The level $k_1$ can be estimated from the last row in the Table~1.

In summary, we investigated the gauge coupling unification in MSSM
with the non-universal soft supersymmetry breaking masses.
We found that in such cases the unification scale could be pulled up 
toward the string unification scale without the large threshold 
correction. 
This seems to be favorable to the superstring unification in the 
duality invariant string models.
The physics of the non-universal soft supersymmetry breaking will 
deserve further investigation for the string unification.

{\bf Acknowledgment}\\
The authors would like to thank Prof. T.~Yanagida for informing us of 
their work\cite{yana} where the similar analyses are done in 
the supersymmetric GUT.
They also thank S.Matsumoto for helpful discussions.

\newpage

{\large\bf Table~1}\\
The change of the soft breaking mass from the low energy region 
($M=M_S$ or $M_Z$) to $M_X$ and the ratio of the 
unification coupling $\alpha_X$ and $U(1)$ coupling $\alpha_1\equiv g_1^2/4\pi$
 at $M_X$. 
For an example we take $M_S=1$~TeV and $M_S=\sqrt 10$~TeV.
$\Delta m_i^2$ is defined as $\Delta m_i^2=m_i^2(M)-m_i^2(M_X)$
and $M_a=M_a(M_Z)$.
The listed values are normalized by the gaugino mass $M_U$ 
at $M_{\rm str}$. 
In the MSSM with $M_S=100$~GeV, $\Delta m_Q^2/M_U^2=6.87$,
$\Delta m_U^2/M_U^2=6.45$,
$\Delta m_L^2/M_U^2=0.53$,
$M_3/M_U=2.86$,
$M_2/M_U=0.82$,
$M_1/M_U=0.40$ and 
$\alpha_X/\alpha_1=1.61$.

\vspace{1cm}

\begin{center}
\begin{tabular}{|c|c|c|c|c|c|c|}\hline
  &{\it Case~I}&{\it Case~II}&{\it Case~III}&{\it Case~IV}
&{\it Case~V}&{\it Case~VI}\\ \hline\hline
$\Delta m_Q^2/M_U^2$ &5.48&4.98&5.31&7.24&4.86&6.79\\
$\Delta m_U^2/M_U^2$ &5.10&4.60&6.74&6.81&4.47&4.64\\
$\Delta m_L^2/M_U^2$ &0.48&0.47&0.50&0.50&0.55&0.50\\ \hline
$ M_3/M_U$ &2.34&3.56&3.25&2.93&3.51&3.18\\
$ M_2/M_U$ &0.84&0.91&0.90&0.85&0.89&0.83\\
$ M_1/M_U$ &0.43&0.42&0.40&0.40&0.42&0.43\\
$ \alpha_X/\alpha_1$&1.57&1.57&1.52&1.57&1.60&1.65\\ \hline
\end{tabular}\vspace{.6cm}\\
$M_S=1$~TeV \vspace{1cm}\\
\begin{tabular}{|c|c|c|c|c|c|c|}\hline
  &{\it Case~I}&{\it Case~II}&{\it Case~III}&{\it Case~IV}
&{\it Case~V}&{\it Case~VI}\\ \hline\hline
$\Delta m_Q^2/M_U^2$ &3.68&4.31&4.68&7.35&4.17&6.84\\
$\Delta m_U^2/M_U^2$ &3.32&3.96&6.97&6.92&3.80&3.98\\
$\Delta m_L^2/M_U^2$ &0.46&0.45&0.49&0.49&0.52&0.48\\ \hline
$ M_3/M_U$ &2.16&3.90&3.43&2.95&3.82&3.31\\
$ M_2/M_U$ &0.85&0.95&0.94&0.86&0.92&0.84\\
$ M_1/M_U$ &0.44&0.43&0.40&0.40&0.43&0.44\\
$ \alpha_X/\alpha_1$&1.55&1.55&1.48&1.55&1.60&1.67\\ \hline
\end{tabular} \vspace{.6cm}\\
$M_S=\sqrt{ 10}$~TeV
\end{center}
\newpage
{\large\bf Figure Caption}\\
{\large\bf Fig.1}\ Unification scale $M_X$ of the gauge couplings of 
$SU(3)$ and $SU(2)$ corresponding to the decoupling scale $M_S$ of 
some superpartners of MSSM. 
Both scales are defined as $M_X=10^y$~GeV and $M_S=10^x$~GeV.
The explanation of each case is given in the text.

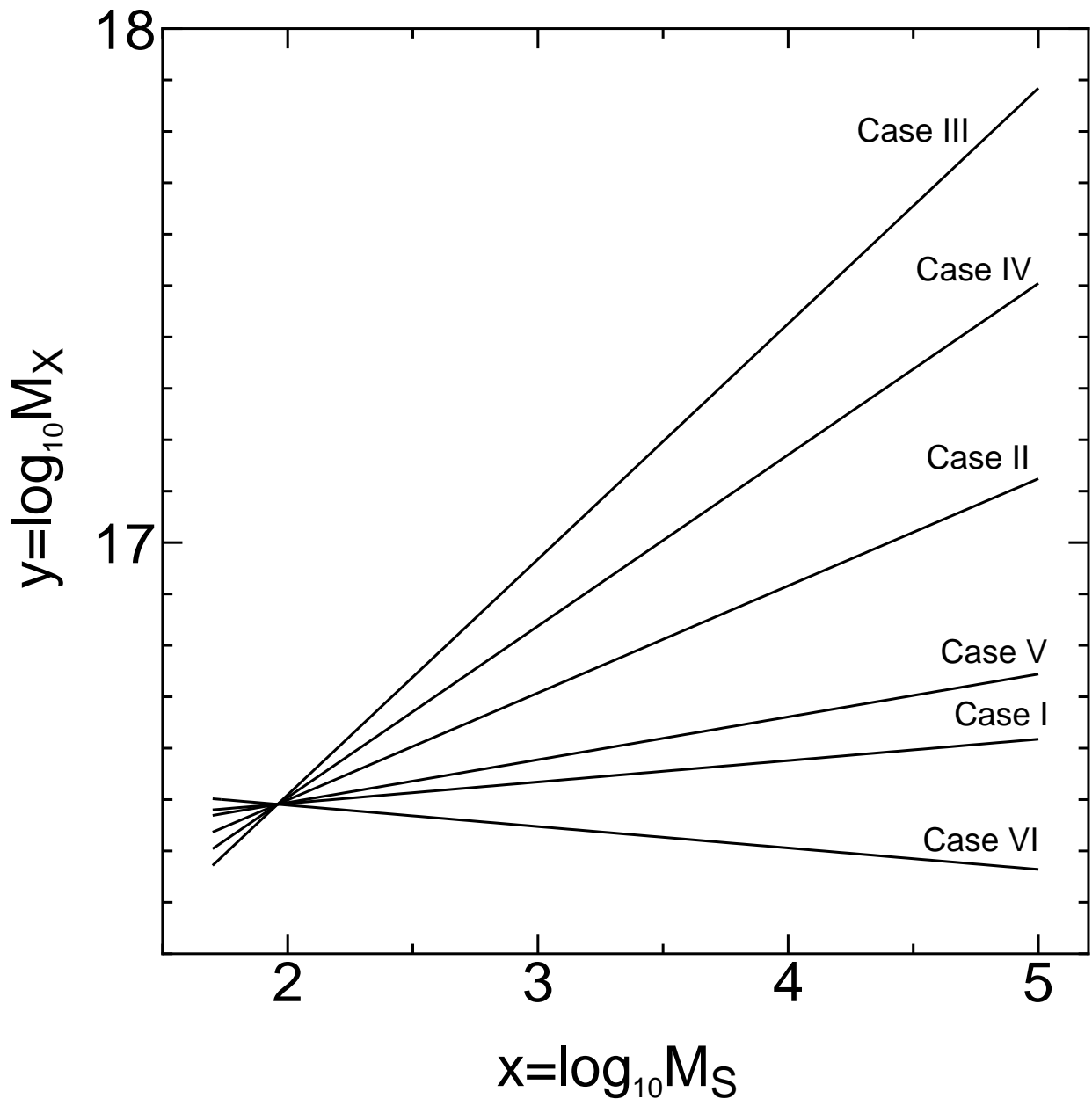
\end{document}